\documentclass[10pt,prd,preprintnumbers,floatfix,aps,nofootinbib,showpacs,twocolumn,superscriptaddress,noshowpacs]{revtex4-1} 
\usepackage{amsthm,amsmath,amssymb,amsfonts}

\usepackage{graphicx}
\usepackage{hyperref}
\usepackage{amsthm,amsmath,amssymb,amsfonts}
\usepackage{amsbsy}
\usepackage{bbm}
\usepackage{mathtools}
\usepackage{microtype}
\usepackage{diagbox}

\usepackage{xcolor}

\newcommand{\be}{\begin{equation}}
\newcommand{\ee}{\end{equation}}

\newcommand{\dd}{\mathrm{d}}

\usepackage{soul}

\begin{document}

\preprint{IFT-UAM/CSIC-24-171, KCL-PH-TH-2024-64}
\title{Search for a gravitational wave background from primordial black hole binaries using data from the first three LIGO-Virgo-KAGRA observing runs}

\author{Tore Boybeyi}
\email{boybe001@umn.edu}
\affiliation{School of Physics and Astronomy, University of Minnesota, Minneapolis, 55455 MN, USA}
\author{Sebastien Clesse}
\email{sebastien.clesse@ulb.be}
\affiliation{Service de Physique Th\'eorique  Brussels Laboratory of the Universe, Universit\'e Libre de Bruxelles, Boulevard du Triomphe, CP225, 1050 Brussels, Belgium}
\affiliation{ Brussels Laboratory of the Universe, Universit\'e Libre de Bruxelles, Boulevard du Triomphe, CP225, 1050 Brussels, Belgium}
\author{Sachiko Kuroyanagi}
    \email{sachiko.kuroyanagi@csic.es}
    \affiliation{Instituto de F\'isica Te\'orica UAM-CSIC, Universidad Aut\'onoma de Madrid, Cantoblanco 28049 Madrid, Spain}
    \affiliation{Department of Physics and Astrophysics, Nagoya University, 464-8602, Nagoya, Japan}
\author{Mairi Sakellariadou}
\email{mairi.sakellariadou@kcl.ac.uk}
\affiliation{Theoretical Particle Physics and Cosmology Group, \, Physics \, Department, \\ King's College London,  University of London,  Strand,  London, WC2R  2LS, UK}

\date{\today}

\begin{abstract}
 Using the cross-correlation data from the first three observing runs of the LIGO-Virgo-KAGRA Collaboration, we search for a gravitational-wave background (GWB) from primordial black holes, 
 arising from the superposition of compact binary coalescence events.
 We consider both early and late binary formation mechanisms 
 and perform Bayesian parameter inference.
From the non-detection of the GWB, we provide constraints on the fraction of primordial black holes contributing to the present dark matter energy density.
\end{abstract}

\maketitle


\section{Introduction} Gravitational-wave (GW) astronomy  has the potential to revolutionize our understanding of the Universe by providing insights into its most energetic and dynamic phenomena. GW detections allow us to explore regions of space-time that are inaccessible through electromagnetic observations.  Since the first binary black hole (BBH) merger detected by the LIGO-Virgo-KAGRA (LVK) collaboration~\cite{LIGOScientific:2016aoc}, the field has rapidly expanded with numerous GW events reported over the past decade~\cite{LIGOScientific:2018mvr,abbott2021gwtc,KAGRA:2021vkt}. These observations have advanced our knowledge of compact objects, their merger mechanisms, and have enabled tests of general relativity in the strong-field regime~\cite{LIGOScientific:2020tif}. Additionally, GW signals have been used to measure the Hubble constant through standard sirens~\cite{LIGOScientific:2021aug}.

Beyond individual events, the gravitational wave background (GWB)—a collective signal from many \textit{unresolved} sources—provides a powerful tool  for probing both astrophysical and cosmological processes. Astrophysical contributions to the GWB arise from unresolved BBH, binary neutron star (BNS), and black hole-neutron star (BHNS) mergers throughout cosmic history~\cite{Regimbau:2011rp}. These compact binaries may form via various channels, including isolated binary evolution and dynamical interactions in dense stellar environments. On the cosmological side, potential GWB sources include phase transitions in the early universe~\cite{Kosowsky:1991ua,Romero:2021kby}, cosmic strings~\cite{Damour:2000wa,LIGOScientific:2021nrg}, and primordial fluctuations from inflation~\cite{Starobinsky:1979ty}. Detecting a cosmological GWB would open a new window into the physics of the early universe and high-energy phenomena beyond the reach of particle accelerators~\cite{Caprini:2018mtu,Sakellariadou:2022tcm}.

In this article, we focus on primordial black holes (PBHs), which are hypothesized to form from the collapse of large density fluctuations in the early universe~\cite{hawking1971gravitationally}. Unlike astrophysical black holes, which are remnants of massive stars, PBHs can span a wide range of masses, from asteroid-like to supermassive scales, depending on the model. PBHs are gaining particular attention as they offer unique insights into conditions in the early universe, such as the spectrum of primordial density fluctuations, phase transitions, and inflationary physics. Moreover, PBHs are considered potential candidates for dark matter~\cite{hawking1971gravitationally,chapline1975cosmological,Carr:2021bzv,green2021primordial} and  for observed compact binary coalescences
(CBCs)~\cite{Bird:2016dcv,Sasaki:2016jop,Clesse:2016vqa}.

Several previous studies have investigated the possibility that some of the detected CBC events originate from PBH mergers~\cite{abbott2020properties,ali2017merger,clesse2018seven,Raidal:2017mfl,Raidal:2018bbj,Hutsi:2020sol,Vaskonen:2019jpv,Chen:2019xse,clesse2022gw190425,de2021gw190521,ando2018primordial,Blinnikov:2016bxu,Hall:2020daa,Wong:2020yig,Jedamzik:2020omx,DeLuca:2021wjr,Liu:2020gif,Vattis:2020iuz,Hall:2020daa,DeLuca:2020qqa,Chen:2024dxhg,Mukherjee:2021ags}. These studies typically analyze the observed population of CBCs 
obtained through individual detections
by examining their mass distributions, spin orientations, and redshift distributions
to discuss
whether the properties of these events are consistent with an astrophysical origin or suggest a primordial origin.
For example, 
a BBH merger
with component masses in the so-called ``mass gap'' region---where standard stellar evolution models predict a scarcity of black holes due to pair-instability supernovae---has prompted speculation about a primordial origin~\cite{abbott2020gw190521,de2021gw190521,Liu:2020gif,de2021gw190521,Franciolini:2021tla,clesse2022gw190425}. 
Furthermore, a subsolar mass event, can serve as distinct evidence of PBHs~\cite{nitz2022broad,Phukon:2021cus,LIGOScientific:2022hai,Miller:2024fpo}.

These analyses compare the observed GW event rates with the predicted merger rates from PBH populations.
Note, however, that LVK binary abundance constraints rely on calculations showing that PBHs form binaries very efficiently in the early universe, treating PBHs as constant Schwarzschild masses. This assumption is not well justified, since during such early times the Hubble horizon is still small. Consequently, one cannot ignore the fact that these PBHs are embedded in an expanding background~\cite{Boehm:2020jwd}.

Another approach is to search
unresolvable binaries as a GWB. Some studies have explored constraints on PBHs by considering the GWB produced by their mergers, a line of research pioneered in \cite{mandic2016stochastic}. 
Constraints have been discussed in the literature~\cite{Raidal:2017mfl,Hutsi:2020sol,wang2018constraints,Inomata:2023zup,Mukherjee:2021itf,Romero-Rodriguez:2024ldc,Atal:2022zux}, typically involving a rough comparison between theoretical predictions and the upper bounds obtained from the isotropic GWB search performed by the LVK collaboration~\cite{abbott2017upper,abbott2019search,abbott2021upper}, which assumes a power-law spectrum $\Omega_{\rm GW} \propto f^{\alpha}$. Although this assumption may be valid for a GWB originating from an unresolved BBH population from Population I or II stars (where lower end is bounded by the Chandrasekhar limit, while the intermediate mass range is constrained by pair-instability), PBH mass distributions can extend significantly beyond $\sim 50\,M_{\odot}$. As a result, the GWB spectrum could deviate from a simple power-law form when PBH sources are considered (\cite{Atal:2022zux}).

Motivated by these, we analyze public GWB data from the first three LVK observing runs (O1--O3)~\cite{abbott2017upper,abbott2019search,abbott2021upper} 
and perform a full Bayesian parameter inference
for the GWB from PBHs with a wide range of masses, including sub-solar masses up to $10^3$ solar masses.
This allows us to account for the complete spectral shape and explore a broad parameter space that extends beyond the detection of individual mergers by LVK. Some recent works in this direction are \cite{Mukherjee:2021ags,Atal:2022zux}, although the former
was primarily focused on distinguishing between astrophysical and primordial BBHs. In addition, the PBH model used in \cite{Mukherjee:2021ags,Atal:2022zux} were limited to a single mass bin of $1$ and $30 M_{\odot}$. It is worth noting that in what follows it is for the first time considered the combination of different formation channels to get corresponding constraints and assess their relative significance. 

In what follows, we aim to effectively bridge the gap between theoretical studies and observational data through robust Bayesian analysis. Updating to the most up-to-date publicly available data from the LVK (O1-O3) searches, we provide more reliable and comprehensive constraints on PBH abundance for peaked mass distributions across a broad mass range, a topic gaining significant attention in the field, particularly in relation to other cosmological constraints. The Bayesian framework offers key advantages in this context. It is a natural framework to include various constraints between parameters an uncertainties on them into the analysis. Moreover, in the event of a future GWB detection, Bayesian inference will be essential for identifying the underlying source population and distinguishing between astrophysical and primordial origins. 

The paper is organized as follows. In Sec.~\ref{sec:model}, we describe the models for calculation of the PBH merger rate. In Sec.~\ref{sec:Bayesian}, we outline our Bayesian inference method and present the results. Then we conclude in Sec.~\ref{sec:conclusion}. Appendix~\hyperref[sec:appd]{A} provides additional details on the suppression factors relevant to the early binary formation scenario. Appendix\hyperref[sec:appd]{B} describes the cross-correlation statistics used in our analysis. Appendix~\hyperref[sec:appd]{C} presents an injection study exploring various potential scenarios. Appendix~\hyperref[sec:appd]{D} provides the posterior distributions of all parameters corresponding to the main results.


\section{Modeling}\label{sec:model}
To parameterize the abundance of PBH, we commonly use $f_{\rm PBH}= \Omega_{\rm PBH}/\Omega_{\rm DM}$, which is the fraction of PBH in the dark-matter energy density today. One of the key factors influencing the GW spectrum in our model is the mass function of PBHs. It is commonly parametrized using monochromatic or Log-Normal distributions. They may not accommodate a mass function with a detailed shape, such as the one associated with the QCD phase transition~\cite{{Jedamzik:1996mr,Carr:2019kxo,Carr:2019hud}}, but it is a good approximation for PBHs arising from a peak in the primordial power spectrum~\cite{Gow:2020cou}.

Using the PBH energy density $\rho_{\rm PBH}$, we define the PBH mass function $p(m)$ as 
\be
p(m) = \frac{1}{\rho_{\rm PBH}}
\frac{\dd \rho_{\rm PBH}}{\dd \ln m}.
\ee
For a Log-Normal Gaussian distribution of the PBH mass function, we have  
\be\label{massfunc}
p(m)=\frac{1}{\sqrt{2\pi}\sigma} \exp\left[-\frac{(\ln m - \ln\mu)^2}{2\sigma^2}\right].
\ee
The mass function is normalized to satisfy $\int p(m) \dd \ln m = 1$ and is usually motivated by inflationary scenarios \cite{PhysRevD.47.4244} . We concentrate on relatively narrow mass distributions ($\sigma < 1$) because theoretical predictions for broader distributions contain significant uncertainties \cite{DeLuca:2020ioi,Gow:2020cou,Ferrante:2022mui,Gow:2022jfb}. Nevertheless, our chosen distribution remains sufficiently broad within the LVK sensitivity range, allowing mass variations on the order of the central value ($\sim\mathcal{O}(\mu)$), potentially spanning several tens of solar masses around $\sim 50 \,M_{\odot}$. While our simple log-normal distribution represents progress towards understanding these mass distributions, detailed modeling incorporating more complex structure will be deferred to future work.

The amplitude of the GWB is often characterized by the dimensionless parameter 
\be\Omega_{\rm GW}\equiv \frac{1}{\rho_{c,0}}\frac{\dd\rho_{\rm GW}}{\dd\ln f},
\ee
where $\rho_{c, 0}=3H_0^2/(8\pi G)$ is the critical density of the Universe today. The GWB formed by the ensemble of CBC events can be calculated by summing up the energy spectrum of each binary system, divided by $(1+z)$ to account for the redshifting of the gravitons since emission and taking into account the merger rate distribution~\cite{Phinney:2001di}
\begin{align}
  \Omega_{\rm GW}(f)&=\frac{f}{\rho_{c,0}}\int_0^{z_{\rm max}}\dd z \int  \dd \ln m_1 \ \dd \ln m_2 \ \frac{p(m_1) p(m_2)}{(1+z)H(z)}\nonumber \\ & \times \frac{d^2 R_{\rm EB/LB}}{d\ln m_1 d\ln m_2} \frac{\dd E_{\rm GW}}{\dd  f_{\rm r}},
  \label{eq:OGW}
\end{align}
where $f_{\rm r}=(1+z)f$ is the GW frequency in the source frame. Here $\frac{d^2 R_{\rm EB/LB}}{d\ln m_1 d\ln m_2}$ describes the differential merger rate per unit
time, comoving volume, and mass interval where EB and LB stand for “Early Binary” and “Late Binary” respectively and denote different binary formation
mechanisms described in the next subsection. $\dd E_{\rm GW}/\dd  f_r$ is the single source energy spectrum, which is a function of the masses of the compact objects, $m_1$ and $m_2$; its 
fitting function including inspiral, ringdown, and merger phases
is given in \cite{ajith2011inspiral}.

\subsection{Early binary formation}\label{sec:suppresion}
 \emph{Early binaries} are formed in the radiation-dominated era shortly after PBH formation. In this scenario, 
a binary is formed from a pair of
closely spaced PBHs with the surrounding third object acting on the pair via a tidal
force that generates the angular momentum of the binary~\cite{Nakamura:1997sm,Sasaki:2016jop}.
The merging rate (per unit logarithmic mass of the two binary components) of early binaries at a time $t$ is given by~\cite{ioka1998black,Raidal:2018bbj,ali2017merger,Sasaki:2016jop,kocsis2018hidden}.
\begin{align}
    \frac{d^2 R_{\rm EB}}{d\ln m_1 d\ln m_2} &= \frac{1.6 \times 10^6}{\rm Gpc^3 yr}  f_{\rm PBH}^{53/37} \Big[\frac{t}{t_0}\Big]^{-34/37} 
    \nonumber \\ & \times\left(\frac{m_1 + m_2}{M_\odot}\right)^{-32/37}    \left[\frac{m_1 m_2}{(m_1+m_2)^2}\right]^{-34/37} \nonumber \\ & \times S(m_1,m_2,f_{\rm PBH}) ,  \label{eq:cosmomerg}
\end{align}
where $m_1$ and $m_2$ are the two black hole masses and $t_0$ is the age of the Universe. A rate suppression factor, $S$, has been introduced to account for two principal mechanisms that impede binary formation due to local matter inhomogeneities and nearby PBHs $S_1(m_1,m_2,f_{\rm PBH})$ and 
the clustering due to their initial
Poissonian fluctuations 
$S_2(f_{\rm PBH})$ \cite{Raidal:2018bbj,Hutsi:2020sol}. Analytical prescriptions for these suppression factors have been studied in \cite{Hutsi:2020sol, Raidal:2018bbj} which are shown to be consistent with numerical simulations \cite{Raidal:2018bbj}. We detail these aspects in Appendix \hyperref[sec:appA]{A}, including the relevant ranges for our work and their limitations.

\subsection{Late binary formation}
\emph{Late binaries} form much later 
due to dynamical capture in the clusters of
PBHs during the matter-dominated era~\cite{Bird:2016dcv}. For a general mass distribution of PBHs, an expression for the merger rate has been proposed in the literature~\cite{1989ApJ...343..725Q,Mouri:2002mc}:
\be 
\frac{d^2 R_{\rm LB}}{d\ln m_1 d\ln m_2} = \frac{R_{\rm clust}}{\rm Gpc^3 yr} f_{\rm PBH}^2  \frac{(m_1 + m_2)^{10/7}}{(m_1 m_2)^{5/7}} \,.
\label{eq
} \ee
$R_{\rm clust}$ is a single parameter that encapsulates how much the local environment of PBHs boosts their merger rate which depends on the velocity dispersion and density contrast in PBH. In $\Lambda \rm CDM$ model without significant small-scale Poisson fluctuations, one typically finds low values in the range of a $O(1\sim 10)$ \cite{PhysRevLett.116.201301} by applying the Press-Schechter formalism to the linear matter power spectrum.  However, if PBHs make up most of the dark matter at around solar masses or above, their discrete (Poisson-like) distribution can seed dense subhalos that raise this factor into the $O(10^2 \sim 10^3)$ \cite{clesse2022gw190425,bagui2022boosted}. Current GW data from individually detected binaries, particularly from mergers near the neutron-star and pair-instability mass gaps, is consistent with $\sim 4 \times 10^2$ \cite{clesse2022gw190425}. This value remains somewhat model-dependent, since processes like direct cluster formation or varying PBH mass fractions can push it higher or lower, but the data suggest that a few hundred is a reasonable benchmark. To encapsulate these different cases, we will take three reference values for $R_{\rm clust}=[1,4 \times 10^2,10^3]$.

\section{Bayesian search}
\label{sec:Bayesian}
\subsection{Methodology}
Let us outline the Bayesian framework employed in our search for GWB originating from PBHs and other compact binary sources. Our methodology was first introduced in \cite{mandic2012parameter} and closely follows that of previous studies~\cite{martinovic2021searching,romero2022search,badger2023probing,badger2024detection,renzini2023pygwb,wu2013accessibility,crowder2013measurement,crocker2015model,fitz2018multiwavelength,abbott2019search,abbott2021upper}, with specific adaptations to our analysis.

We utilize data from the Advanced LIGO detectors located in Hanford (H) and Livingston (L), as well as data from the Virgo (V) observatory, during the first three observing runs (O1–O3) \cite{PhysRevLett.123.231108,PhysRevLett.116.131103,PhysRevD.102.062003,KAGRA:2023pio}. Our search incorporates all available baselines (HL, HV, LV) (only HL baseline is available for O1 and O2) to maximize sensitivity and leverage the complementary geographical locations of the detectors. 

Following the notation of~\cite{renzini2023pygwb}, 
the primary observable is the cross-correlation estimator \( \hat{C}_{IJ,f_i} \) which measures the correlation between detectors $I$ and $J$ as a function of discrete frequency $f_i$. This estimator is constructed by cross-correlating the discrete Fourier transforms of the outputs from different detectors. It is then transformed into the estimator $\hat{\Omega}^{IJ}_{{\rm GW},f_i}$ by multiplying with the relevant numerical factors and the overlap reduction function. These definitions are given in Appendix \hyperref[sec:appb]{B}. 
The likelihood function for the search of an isotropic
GWB is~\cite{mandic2012parameter,renzini2023pygwb}
\begin{equation}\label{eq:likelihood}
p(\hat{\Omega}^{IJ}_{{\rm GW},f_i} | \boldsymbol{\theta}, \lambda) \propto \exp \left[ -\frac{1}{2}\sum_{IJ}\sum_{f_i} \left( \frac{ \hat{\Omega}^{IJ}_{{\rm GW},f_i} - \lambda \Omega_{\rm GW}(f_i | \boldsymbol{\theta}) }{\sigma^{IJ}_{f_i}} \right)^2 \right] \,,
\end{equation}
where \( \sigma^{IJ}_{f_i} \) denotes the variance of the cross-correlation estimator, which is derived from the noise power spectral density of each detector and $\Omega_{\rm GW}(f_i | \boldsymbol{\theta})$ represents the theoretical model to be compared with the data, characterized by a set of parameters $\boldsymbol{\theta}$. The sum runs over the frequency bins \( f_i \) as well as different pairs of the detectors labeled by $I, J = {\rm H, L, V}$. We then construct the posterior distribution using Bayes' Theorem,
\begin{align}
   p(\boldsymbol{\theta}, \lambda|\hat{\Omega}^{IJ}_{{\rm GW},f_i} ) =  \frac{p(\hat{\Omega}^{IJ}_{{\rm GW},f_i} | \boldsymbol{\theta}, \lambda)  p (\theta,\lambda)}{p(\hat{\Omega}^{IJ}_{{\rm GW},f_i} )} \,.
\end{align}
We then marginalize over the calibration uncertainty parameter \( \lambda \),  
as detailed in calibration uncertainty studies~\cite{whelan2014treatment,sun2020characterization,renzini2023pygwb}, to get $p(\boldsymbol{\theta}|\hat{\Omega}^{IJ}_{{\rm GW},f_i} ) = \int d\lambda \ p(\lambda) p(\boldsymbol{\theta}, \lambda|\hat{\Omega}^{IJ}_{{\rm GW},f_i} ) $.

\subsection{Prior Distributions} 
We employ a set of priors includes the parameters: the mean of the Log-Normal mass distribution \( \mu \), its width \( \sigma \), and the PBH fraction \( f_{\rm PBH} \) in the dark matter energy density today, as introduced previously. 


\begin{figure*}[ht]
    \centering
    \includegraphics[width=1.\linewidth]{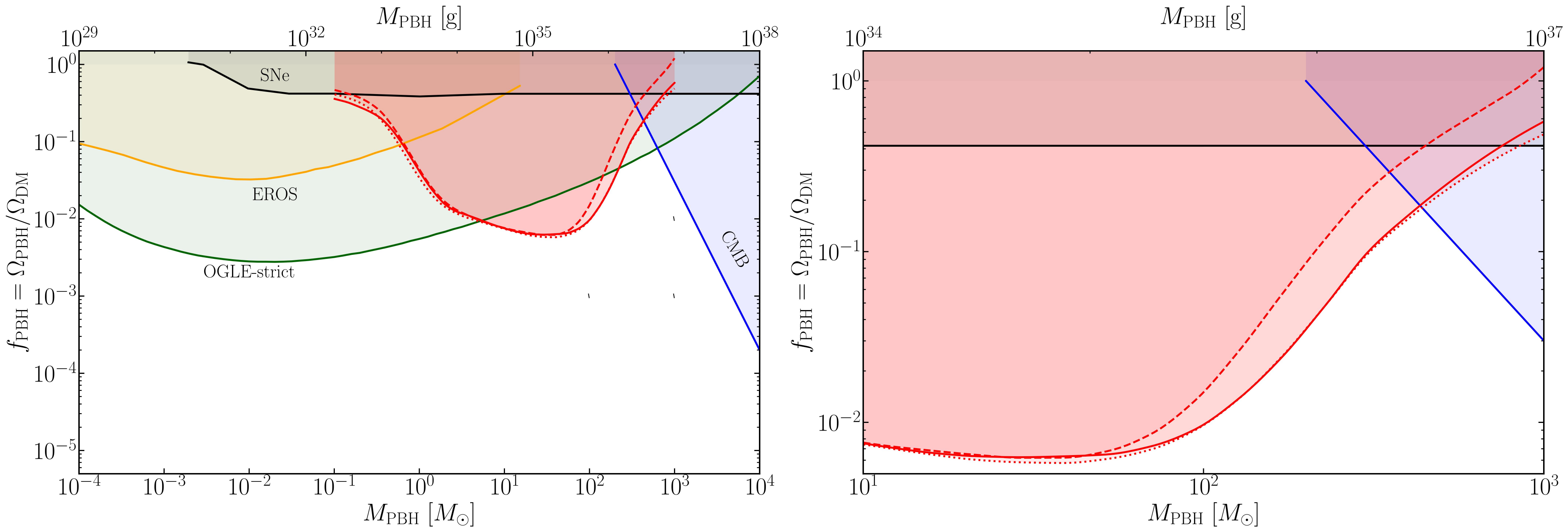}
    \caption{(Left) 95\% C.L. constraints on $f_{\rm PBH}$  as a function of $M_{\rm PBH}$($=\mu$) from the search for the GWB using the different $R_{\rm clust}$ choices are shown in red ($1$-dashed, $4 \times 10^2$-solid, $10^3$-dotted). We also display supernova lensing constraints (SNe)~\cite{zumalacarregui2018limits} in purple, Massive Compact Halo Object (MACHO) constraints~\cite{tisserand2007limits} in orange designated as (EROS), the recent Optical Gravitational Lensing Experiment (OGLE) constraints are shown in green with different Milky Way (MW) models~\cite{mroz2024no}, and cosmic microwave background (CMB) constraints from~\cite{Agius:2024ecw}. We used the \textsc{PBHBounds} \cite{bradley_j_kavanagh_2019_3538999} to generate this plot. (Right) Zoomed‑in view of the same plot.}
    \label{fig:constraints}
\end{figure*} 

We also take into account for an additional otherwise individually unresolvable CBC background 
from astrophysical black holes which can include lighter binaries like neutron stars as well. Their spectrum is 
well approximated by a simple fiducial power law characterized by a single parameter, and is expressed as
$\Omega_{\rm cbc}$ as $\Omega_{\rm CBC}(f) = \Omega_{\rm cbc} (f/f_{\rm ref})^{2/3} $, with $f_{\rm ref}=25$~Hz~\cite{callister2016limits,abbott2017upper,abbott2019search,abbott2021upper}. Deviations from this fiducial power law are not prominent and the expected CBC spectrum is only deviates from the power law beyond the current sensitivity (e.g. Fig. 5 in \cite{abbott2021upper}).
Therefore, our model is $\Omega_{\rm GW} = \Omega_{\rm CBC} + \Omega_{\rm EB} + \Omega_{\rm LB} $.

The prior distributions for the model parameters are specified in Table ~\ref{tab:priors}. These ranges encompass plausible values based on theoretical models and existing constraints~\cite{Raidal:2017mfl,Hutsi:2020sol,abbott2021upper,clesse2022gw190425,bagui2022boosted,Hall:2020daa,DeLuca:2021wjr,Chen:2024dxhg,Franciolini:2021tla}. We adopt a log‐uniform prior for the PBH abundance $f_{\rm PBH}$ because a Jeffreys (log‐uniform) prior is the least informative choice for a scale parameter—it assigns equal weight to each decade  avoids biasing the inference toward any particular order of magnitude \cite{Trotta:2008qt}. We set the lower bound on $f_{\rm PBH}$ prior to $10^{-5}$ which is usually the standard choice \cite{Franciolini:2021tla,Andres-Carcasona:2024wqk}, this makes our results directly comparable to the existing literature.

\subsection{Result}
In order to perform Bayesian parameter inference,
we employ the nested sampling algorithm from the \texttt{dynesty} package~\cite{speagle2020dynesty} for robust parameter space exploration.  To expedite the computation of integrals in the GW spectrum calculations, we utilize Sobol Monte Carlo integration~\cite{caflisch1998monte}, a quasi-random low-discrepancy sequence method that improves convergence rates over traditional Monte
Carlo approaches, thereby reducing computational overhead. To account for calibration uncertainties in GW detector data, we use the Python library \texttt{pygwb}~\cite{renzini2023pygwb}, which integrates these uncertainties into the parameter estimation process.

\begin{table}
    \centering
    \begin{tabular}{|l|l|}
        \hline
        \textbf{Parameter} & \textbf{ Prior} \\
        \hline
        $\Omega_{\rm CBC}$ &  $\mathcal{LU}$($10^{-10}$, $10^{-7}$)  \\
        $\mu \, (M_{\odot})$ &  $\mathcal{LU}$($10^{-1}$, $10^{3}$) \\
        $\sigma $ &  $\mathcal{LU}$($10^{-2}$, $10^{0}$)  \\
        $f_{\rm PBH}$ &  $\mathcal{LU}$($10^{-5}$, $1$) \\
        \hline
    \end{tabular}
    \caption{Prior distributions for the model parameters in  in the Bayesian analysis. $\mathcal{LU}$ denotes a LogUniform distribution.}
    \label{tab:priors}
\end{table}

Our analysis is based on the combined cross-correlation dataset from the first three observing runs of the LVK collaboration. We applied data quality cuts and vetoes following the procedures outlined in previous isotropic GWB searches by the LVK~\cite{abbott2017upper,abbott2019search,abbott2021upper}. For parameter sampling, we utilized $5 \times 10^4$ live points with a convergence criterion of $d\ln(z) < 0.1$. 

The results indicate that no significant GWB associated with PBHs or CBCs has been detected over the noise hypothesis, with logarithmic Bayes factors $\sim -1 $ and corner plots in Fig. ~\ref{fig:corner1},~\ref{fig:corner2},~\ref{fig:corner3}. We establish upper limits on the CBC energy density parameter, $\Omega_{\rm cbc}$, of $\sim 3 \times 10^{-9}$ at the $95\%$ C.L.\ level which are consistent with LVK constraints~\cite{abbott2021upper}. 

From the non-detection, we can constrain the fraction of PBHs, $f_{\rm PBH}$, as a function of the mean mass $\mu$. The obtained constraint, along with other studies from lensing surveys like EROS (\cite{tisserand2007limits}), OGLE (\cite{mroz2024no}), supernova Ia (\cite{zumalacarregui2018limits}) and cosmic microwave background (CMB) constraints for different $R_{\rm clust}$ choices are summarized in Fig.~\ref{fig:constraints} and Table \ref{tab:PBH-abundance}. We also provide the resulting corner plots for all model parameters in Appendix \hyperref[sec:appd]{D}. 

The constraint on $f_{\rm PBH}$ is obtained by marginalizing over the mass function width $\sigma$ and $\Omega_{\rm CBC}$.

For $M_{\rm PBH} \gtrsim 3 \times 10^{2}M_{\odot}$, the GWB spectral amplitude is dominated by the late binary formation channel. Although the LVK loses its constraining power sharply in this mass range due to the limited frequency range, this suggests an interesting possibility for constraining this mass range using the late binary formation channel with the Einstein Telescope \cite{Punturo:2010zz,hild2011sensitivity} and Cosmic Explorer\cite{abbott2017exploring,Reitze:2019iox}, which have greater sensitivity at low frequencies. We also demonstrate that Advanced LIGO A+ can provide valuable information about the underlying background. See Appendix \hyperref[sec:appd]{C} for one such scenario.

These findings demonstrate that GWB constraints are and will continue to be a vital probe, particularly within the mass range $[10^{1}, 3\times 10^{2}] \, M_{\odot}$. Our results improve the previous GWB constraints \cite{Hutsi:2020sol,DeLuca:2020qqa,wang2018constraints} by $  O(2\sim 10)$ at $100 \, M_{\odot}$ with the latest publicly available LVK observing run data. 

\begin{table}[htbp]
  \centering
  \caption{95\% C.L.\ on $f_{\rm PBH}$ at various masses.}
  \label{tab:PBH-abundance}
  \begin{tabular}{|c|c|c|c|}
    \hline
    \diagbox{$R_{\rm clust}$}{$\mu$}
      & $1\,[M_\odot]$     & $30\,[M_\odot]$    & $10^3\,[M_\odot]$   \\
    \hline
       $1$ & $3.6 \times 10^{-2}$  & $5.3 \times 10^{-3}$  & $1.0 \times 10^{0}$ \\
           \hline
     $4\times10^2$           & $4.0 \times 10^{-2}$  & $6.2 \times 10^{-3}$  & $5.8 \times 10^{-1}$\\
    \hline
    $10^3$        & $3.3 \times 10^{-2}$  & $5.8 \times 10^{-3}$  & $4.8 \times 10^{-1}$ \\
    \hline
  \end{tabular}
\end{table}

\section{Conclusion}
\label{sec:conclusion}
In this study, we conducted a targeted search for a GWB originating from PBHs using cross-correlation data from the first three observing runs of the LVK Collaboration~\cite{abbott2017upper,abbott2019search,abbott2021upper}. Unlike previous research that focused on the population of resolved GW events,
we utilized cross-correlated LVK data specifically to search for a GWB resulting from the superposition of many unresolved binary events. Our mass range encompasses PBH binaries with component masses both below and above those detected in the LVK catalog~\cite{LIGOScientific:2018mvr,KAGRA:2021vkt}.

Our approach advances beyond previous stochastic analyses that 
utilized the LVK upper limit derived under the assumption of a power-law shape for the spectrum.
By directly modeling the expected GWB signal from PBH mergers, we present, for the first time, a direct GWB search specifically targeting 
PBH mergers.

Although we did not find significant evidence for a GWB from PBHs, we established new constraints on the fraction of PBHs contributing to the current dark matter energy density, $f_{\rm PBH}$, as a function of PBH mass assuming a log-normal mass distribution. These constraints complement existing bounds from individual CBC events~\cite{Raidal:2017mfl,Hutsi:2020sol,Hall:2020daa} and other observational methods such as microlensing and cosmic microwave background observations~\cite{zumalacarregui2018limits,tisserand2007limits,mroz2024no,Agius:2024ecw}. 


Future constraints will undoubtedly provide deeper and broader insights into the abundance of PBHs across a wider mass range and will offer valuable information for pinpointing the origin of black holes. Additionally, the broad mass function, in particular those predicted by QCD phase transitions, remains an interesting case for investigation. In future work, we plan to apply our analysis to this model using more accurate descriptions of their merger rates that are currently under development. 

\section*{Acknowledgments}
The authors would like to thank Vuk Mandic and the LVK Stochastic group for comments. 

This research has made use of data or software obtained from the Gravitational Wave Open Science Center (gwosc.org), a service of the LIGO Scientific Collaboration, the Virgo Collaboration, and KAGRA. This material is based upon work supported by LIGO Laboratory which is a major facility fully funded by the National Science Foundation, as well as the Science and Technology Facilities Council (STFC) of the United Kingdom, the Max-Planck-Society (MPS), and the State of Niedersachsen/Germany for support of the construction of Advanced LIGO and construction and operation of the GEO600 detector. Additional support for Advanced LIGO was provided by the Australian Research Council. Virgo is funded, through the European Gravitational Observatory (EGO), by the French Centre National de Recherche Scientifique (CNRS), the Italian Istituto Nazionale di Fisica Nucleare (INFN) and the Dutch Nikhef, with contributions by institutions from Belgium, Germany, Greece, Hungary, Ireland, Japan, Monaco, Poland, Portugal, Spain. KAGRA is supported by Ministry of Education, Culture, Sports, Science and Technology (MEXT), Japan Society for the Promotion of Science (JSPS) in Japan; National Research Foundation (NRF) and Ministry of Science and ICT (MSIT) in Korea; Academia Sinica (AS) and National Science and Technology Council (NSTC) in Taiwan.
The software packages used were {\tt matplotlib}~\cite{hunter2007matplotlib}, {\tt numpy}~\cite{van2011numpy}, {\tt scipy}~\cite{virtanen2020scipy}, {\tt dynesty}~\cite{speagle2020dynesty}, {\tt pygwb}~\cite{renzini2023pygwb} and {\tt PBHBounds}  \cite{bradley_j_kavanagh_2019_3538999}. This manuscript was assigned LIGO-Document number LIGO-P2400579.

SC acknowledges support from the Belgian Fund for Research FNRS-F.R.S. through an Incentive Grant For Scientific Research and through the IISN convention 4.4501.19.

SK is supported by the Spanish Attraccion de Talento contract no. 2023-5A-TIC-28945 granted by the Comunidad de Madrid, the I+D grant PID2023-149018NB-C42 and the Grant IFT Centro de Excelencia Severo Ochoa No CEX2020-001007-S funded by MCIN/AEI/10.13039/501100011033, the Consolidaci\'on Investigadora 2022 grant CNS2022-135211, and Japan Society for the Promotion of Science (JSPS) KAKENHI Grant no. JP20H05853, and JP23H00110, JP24K00624.

MS acknowledges support from the Science and Technology Facility Council
(STFC), UK, under the research grant ST/X000753/1.

\bibliography{main}

\clearpage


\section*{Appendix A: Models for Suppression Factors}
\label{sec:appA}
Here we give for completeness the models for binary suppression factors for early PBH binaries that are studied in \cite{Raidal:2018bbj,Hutsi:2020sol}. 

The suppression factor is expressed as the product of $S_1$ and $S_2$. The first factor, $S_1$, describes the disruption of binaries due to either matter fluctuations or the influence of a nearby PBH. It is given by
\begin{eqnarray}
S_1 \approx 1.42 \left[ \frac{(\langle m_{\rm PBH}^2 \rangle/\langle m_{\rm PBH} \rangle^2)}{\bar N + C} + \frac{\sigma_{\rm M}^2}{f_{\rm PBH}^2}\right]^{-21/74} {\rm e}^{-\bar N},
\label{eq:S1}
\end{eqnarray}
where $\sigma_M = 0.005$ is the variance of matter density perturbations, evaluated at binary formation, and $\bar N$ is the number of PBHs within a sphere of comoving radius around the initial PBH pair, given by
\begin{eqnarray}
\bar N = \frac{m_1+m_2}{\langle m_{\rm PBH} \rangle} \frac{f_{\rm PBH}}{f_{\rm PBH}+\sigma_M}~,  \label{eq:Nbar}
\end{eqnarray}
In Eqs.~(\ref{eq:S1}) and~(\ref{eq:Nbar}), the mean PBH mass and their variance are calculated from the mass function through
\begin{align}\label{int1}
\langle m_{\rm PBH} \rangle & =   \frac{\int m_{\rm PBH} \dd n_{\rm PBH}}{n_{\rm PBH}} \nonumber \\ &=  \left[\int \frac{f(m_{\rm PBH})}{m_{\rm PBH}} \dd \ln m_{\rm PBH} \right]^{-1} \\ 
\langle m_{\rm PBH}^2 \rangle &= \frac{\int m_{\rm PBH}^2 \dd n_{\rm PBH}}{n_{\rm PBH}}
 \nonumber\\&=  \frac{\int m_{\rm PBH} f(m_{\rm PBH}) \dd \ln m_{\rm PBH}}{\int \frac{f(m_{\rm PBH}) }{m_{\rm PBH}}\dd \ln m_{\rm PBH}}~, \label{int2}
\end{align}
where $n_{\rm PBH}$ denotes the total PBH number density.  The function $C$ encodes the transition between small and large $\bar N$ limits.  A good approximation is
\begin{align}
C &\simeq \frac{f_{\rm PBH}^2 \langle m_{\rm PBH}^2 \rangle}{\sigma_{\rm M}^2 \langle m_{\rm PBH} \rangle^2}  
\nonumber\\&\times \left\{ \left[ \frac{\Gamma(29/37)}{\sqrt{\pi} } U\left( \frac{21}{74},\frac{1}{2} , 
\frac{5 f_{\rm PBH}^2}
{6 \sigma_{\rm M}^2}
\right) 
\right]^{-74/21}  -1 \right\}^{-1},
\end{align}
where $\Gamma$ is the Euler function and $U$ is the confluent hypergeometric function.

\begin{figure}[h]
    \centering
    \includegraphics[width=\linewidth]{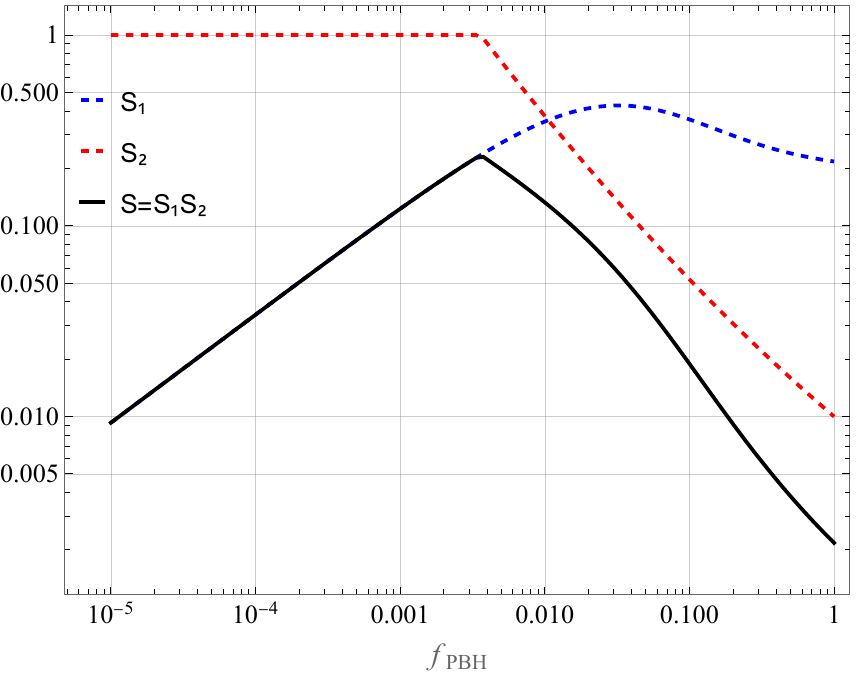}
    \caption{Binary suppression factors for early binary formation are presented as a function of $f_{\rm PBH}$ for a narrow mass distribution. See also Fig. 20 in \cite{LISACosmologyWorkingGroup:2023njw}. }
    \label{fig:sup}
\end{figure}

The second factor, $S_2$, captures the impact of binary disruption within PBH clusters that can form rapidly following PBH formation, provided their abundance is sufficiently high, as a result of initial Poisson fluctuations. It can be approximated by 
\begin{equation}
S_2 \approx \min \left(1,9.6 \times 10^{-3} f_{\rm PBH}^{-0.65} {\rm e}^{0.03 \ln^2 f_{\rm PBH}} \right).
\label{eq:S2}
\end{equation}
We plot 
$S_1$ and $S_2$
in Fig.~\ref{fig:sup} as a function of $f_{\rm PBH}$
for a narrow mass distribution, in which case the above a functions have simpler forms, since we can approximate $\langle m_{\rm PBH}^2 \rangle/\langle m_{\rm PBH} \rangle^2 \approx 1$ and $m_1+m_2/\langle m_{\rm PBH} \rangle \approx 2$. Note that the validity of the above expressions for a broad mass distribution necessitates detailed analytical investigation or validation through N-body simulations. When considering a wide range of masses, the system exhibits a richer phenomenology, including, for example, the presence of PBHs that stimulate cluster formation, as well as smaller PBHs that could dominate the calculation of $\bar{N}$ without being sufficiently massive to disrupt a significant binary system (we refer the reader to Section 4.1.6 of \cite{LISACosmologyWorkingGroup:2023njw} for more details). 
Furthermore, it is important to recognize that early binaries can be significantly affected by their environment throughout cosmic history. This environment undergoes complex evolution driven by factors such as cluster formation following matter-radiation equality, accretion processes, and dynamical heating. As a result, one should exercise caution, as making claims based on these merging rates may be premature. This aspect will be further emphasized in scenarios where PBHs exhibit a broad mass distribution. PBH clusters with varying mass black holes would display a complex array of phenomena, including mass segregation and the ejection of smaller black holes due to kicks, etc.
Currently, we have not yet achieved a comprehensive understanding of such scenarios; therefore, in this study, we concentrate on the case of a narrow mass distribution.

\section*{Appendix B: Cross correlation statistics}\label{sec:appb}
To estimate the GWB, the strain data from each GW detector, denoted by \( s_i(t) \), are divided into segments of duration \( T \). This segmentation helps mitigate the effects of non-stationarities and data interruptions. For each segment \( t_i \) and for every pair of detectors \( I \) and \( J \) (referred to as a baseline), the cross spectral density \( \hat{C}_{IJ,t_i}(f_j) \) is calculated at discrete frequency bins \( f_j \). The cross spectral density is given by
\be
\hat{C}^{IJ}_{t_i,f_j} = \frac{2}{T} \, \text{Re} \left[ s^{*I}_{t_i,f_j} \, s^J_{t_i,f_j} \right],
\ee
where \( s^I_{t_i,f_j} \) represents the Fourier transform of the time series from detector \( I \) within segment \( t_i \) at frequency \( f_j \).

Next, the optimal cross-correlation statistic for GWB, \( \hat{\Omega}^{IJ}_{t_i,f_j} \) and its associated variance \( \hat{\sigma}^{IJ}_{t_i,f_j} \) are constructed to maximize the signal-to-noise ratio (SNR) for each narrow frequency bin \( f_j \). These are obtained as \cite{Allen:1997ad,romano2017detection,renzini2023pygwb}
\be
\hat{\Omega}^{IJ}_{t_i,f_j} = \frac{\hat{C}_{IJ,t_i}(f_j)}{\gamma_{IJ}(f_j) \, S_0(f_j)},
\ee
\be
(\hat{\sigma}^{IJ}_{t_i,f_j})^2 = \frac{1}{2 T \Delta f} \cdot \frac{P_{1,t_i}(f_j) \, P_{2,t_i}(f_j)}{\gamma^2_{IJ}(f_j) \, S^2_0(f_j)},
\ee
where \( \gamma_{IJ}(f_j) \) is the overlap reduction function for the baseline \( IJ \), \( S_0(f) = \frac{10 \pi^2 f^3}{3 H_0^2} \) is the normalization factor with \( H_0 \) being the Hubble constant, and \( \Delta f = f_{j+1} - f_j \) is the width of the frequency bin centered at \( f_j \). Additionally, \( P_{1,t_i}(f_j) \) and \( P_{2,t_i}(f_j) \) are the power spectral densities of detectors \( I \) and \( J \) in segment \( t_i \) at frequency \( f_j \).

Then, different segments can be combined with weights $1/(\hat{\sigma}^{IJ}_{t_i,f_j})^2$ to obtain averaged GWB spectra
\begin{align}
    \hat{\Omega}^{IJ}_{\text{GW},f_j} = \frac{\sum_{i} \hat{\Omega}^{IJ}_{t_i,f_j}(\hat{\sigma}^{IJ}_{t_i,f_j})^{-2}}{\sum_{i} (\hat{\sigma}^{IJ}_{t_i,f_j})^{-2}}\,,
\end{align}
which is being used in constructing the likelihood in Eq. \eqref{eq:likelihood}.

\section*{Appendix C: Injection Study}\label{sec:appc}

In order to validate the Bayesian formalism described in the text and provide a possible interesting scenario, we performed an injection study by adding a synthetic GWB signal
\[
\Omega_{\rm GW}(f) \;=\; \Omega_{\rm EB}(f)\;+\;\Omega_{\rm LB}(f)\;+\;\Omega_{\rm CBC}(f)\,.
\]
We injected our model with the parameter values
\begin{align}
&(\mu,\;\sigma,\;f_{\rm PBH},\;R_{\rm clust},\;\Omega_{\rm CBC})
\; = \nonumber \\ & \;(5 \times 10^2 \, M_{\odot},\; 2 \times 10^{-2},\;2 \times 10^{-1},\;4 \times 10^2,\;8 \times 10^{-10})\,, \nonumber
\end{align}
to the projected Advanced LIGO A+ sensitivity curve. The injected spectrum was chosen such that it lies below the O3 Power Law Integrated (PI) curve but above the A+ sensitivity. 
Recovery with our nested‑sampling pipeline yielded accurate posterior constraints on the primary parameters $\mu$, $f_{\rm PBH}$ and $\Omega_{\rm CBC}$; however, $R_{\rm clust}$ and $\sigma$ were not well recovered. This behavior arises because the spectral cutoff induced by $\sigma$ has only a mild impact on the overall shape, making $\sigma$ 
difficult to constrain, and because the late‑binary component is subdominant, reducing sensitivity to $R_{\rm clust}$. These findings are 
corroborated by the corner plots presented in Appendix \hyperref[sec:appd]{D}, 
which show an essentially flat posterior for $\sigma$, and by the upper‑limit curves in the main text, which exhibit only a weak dependence on $R_{\rm clust}$.

\begin{figure}[h]
  \centering
  \begin{minipage}{0.5\textwidth}
    \centering
    \includegraphics[width=0.9\linewidth]{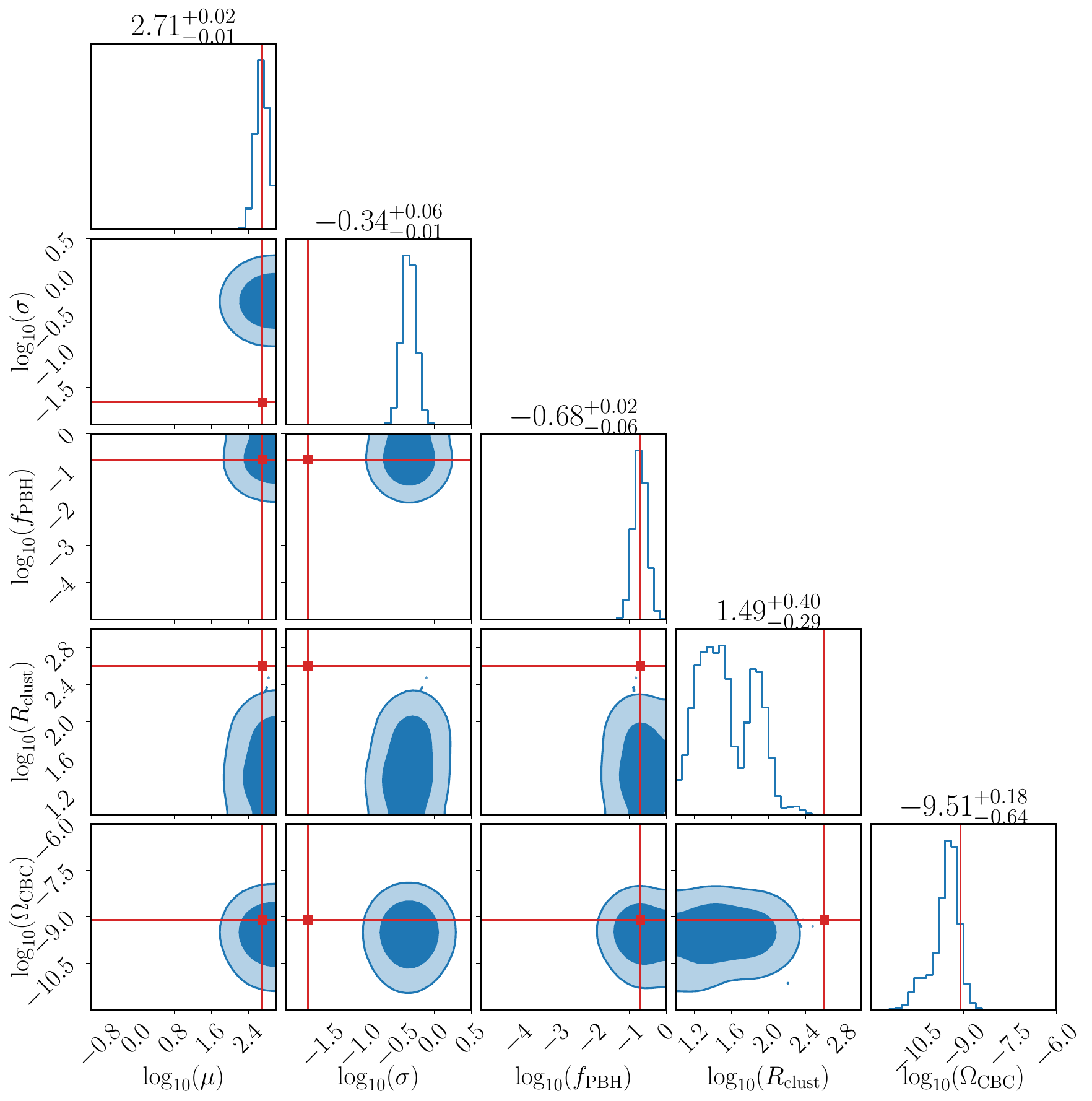}
  \end{minipage}\hfill
  \begin{minipage}{0.5\textwidth}
    \centering
    \includegraphics[width=0.9\linewidth]{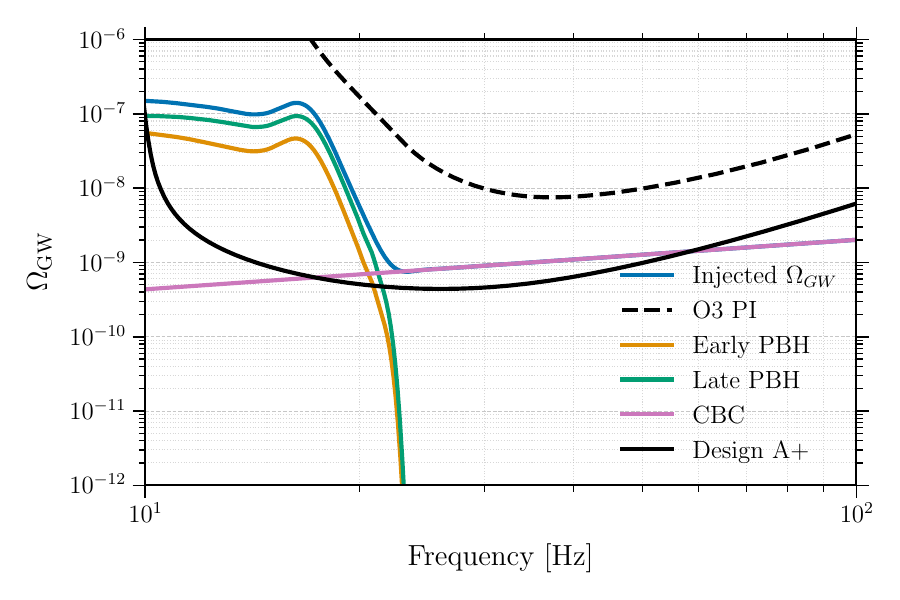}
  \end{minipage}
  \caption{(Left) Posterior distributions from the injection study with true values.
  The diagonal panels show the 1D marginals with the injected values marked by red lines, 
  and the off‐diagonal panels show the 2D joint posteriors with 68\% and 95\% credible contours. (Right) Components of the signal with sensitivity curves. }
    \label{fig:injection}
\end{figure}

\clearpage
\onecolumngrid
\section*{Appendix D: Corner Plots}\label{sec:appd}
\begin{figure}[h]
    \centering
    \includegraphics[width=\linewidth]{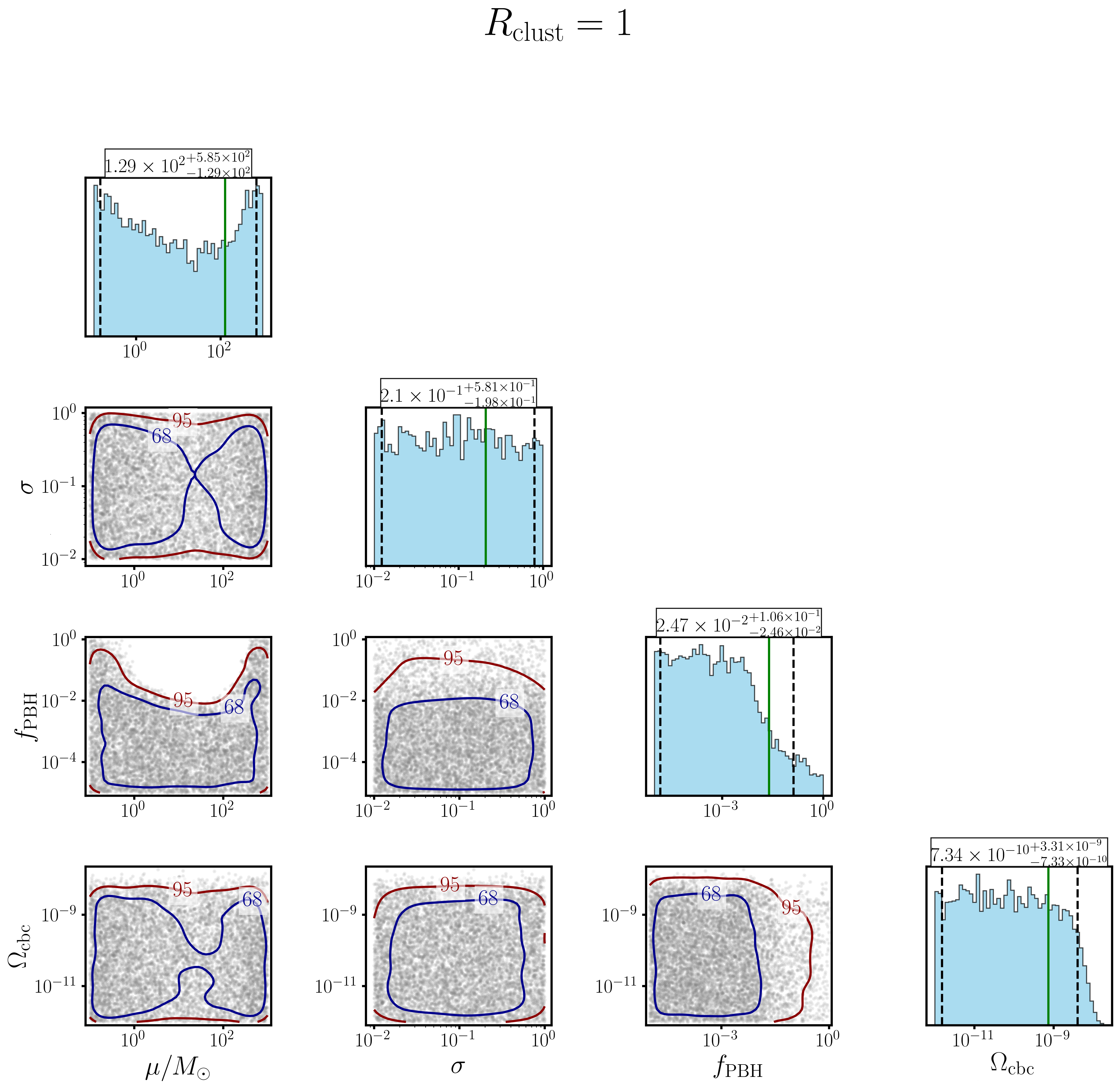}
    \caption{Corner plots of posteriors for $R_{\rm clust}=1$ used in the analysis.}
    \label{fig:corner1}
\end{figure} 
\begin{figure}[h]
    \centering
    \includegraphics[width=\linewidth]{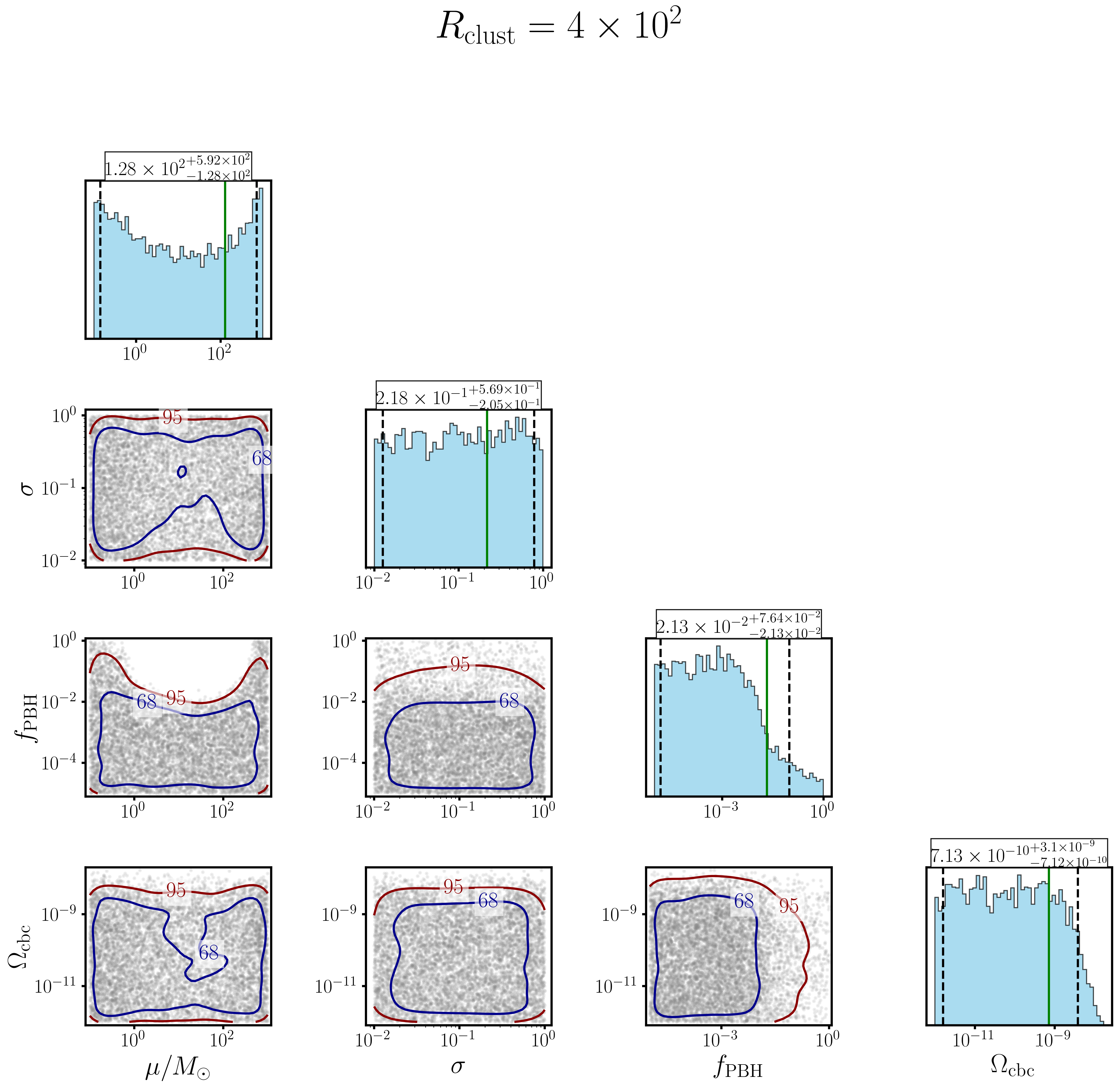}
    \caption{Corner plots of posteriors for $R_{\rm clust}=4 \times 10^2$ used in the analysis.}
    \label{fig:corner2}
\end{figure} 
\begin{figure}[h]
    \centering
    \includegraphics[width=\linewidth]{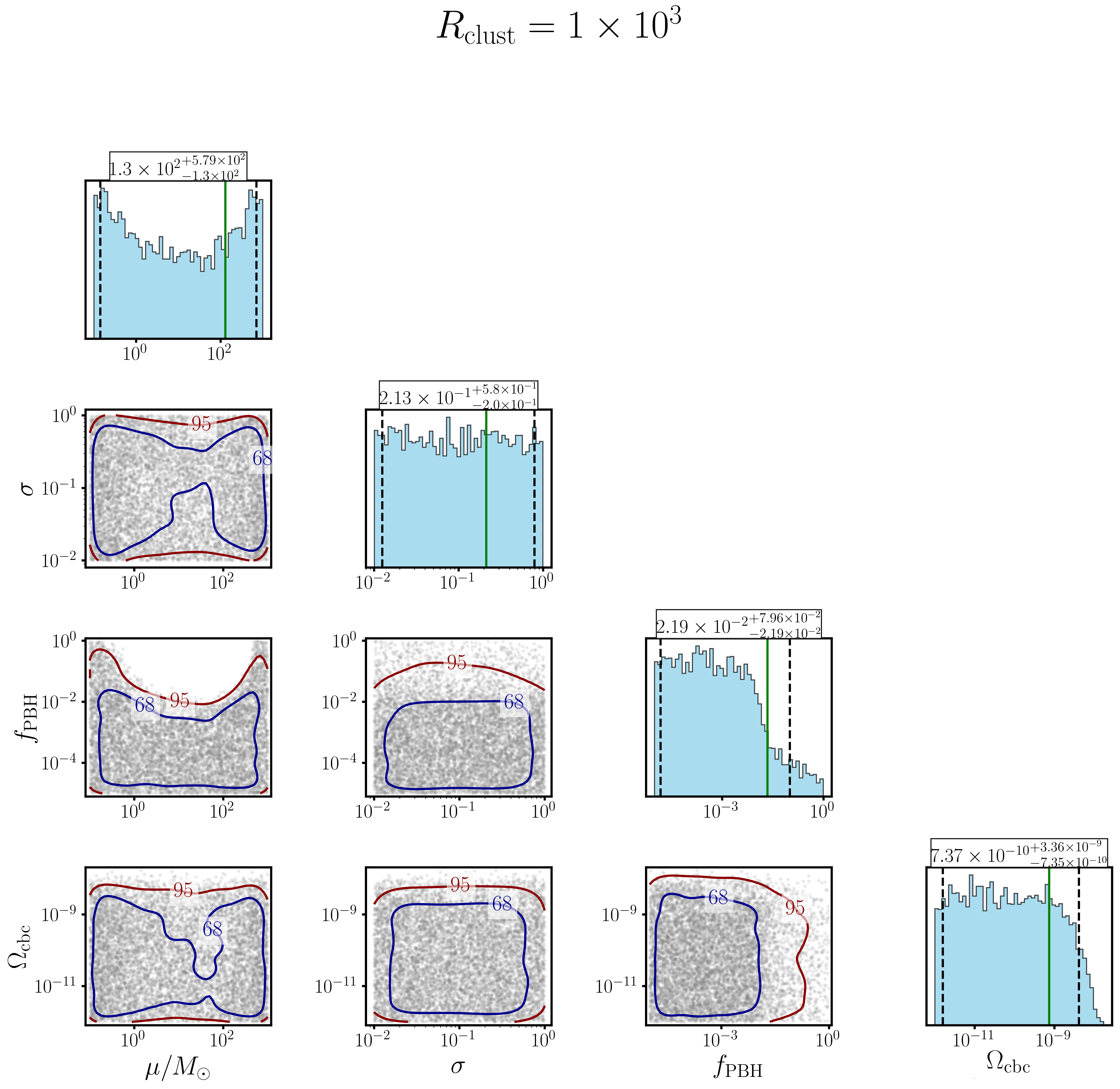}
    \caption{Corner plots of posteriors for  $R_{\rm clust}= 10^3$ used in the analysis.}
    \label{fig:corner3}
\end{figure}



\end{document}